# Study on the influence that the number of positive ion sources has in the propulsion efficiency of an asymmetric capacitor in nitrogen gas

**A A Martins[1] and M J Pinheiro[2]**

[1]Institute for Plasmas and Nuclear Fusion & Instituto Superior Técnico,
Av. Rovisco Pais, 1049-001 Lisboa, Portugal
Email: aam@ist.utl.pt

[2]Department of Physics and Institute for Plasmas and Nuclear Fusion,
Instituto Superior Técnico, Av. Rovisco Pais, 1049-001 Lisboa, Portugal
Email: mpinheiro@ist.utl.pt

**Abstract:** The present work intends to compare the propulsion force developed in an asymmetric capacitor, according to the number of positive ion sources used. The ion source is a corona wire, which generates a positive corona discharge in nitrogen gas directed towards the ground electrode. We are going to apply the known theory of electrohydrodynamics (EHD) and electrostatics in order to compute all hydrodynamic and electrostatic forces that act on the considered geometry in an attempt to provide a physical insight on the force mechanism that acts on each asymmetrical capacitor, and also to compare propulsion efficiencies.

## 1. Introduction

In this work we investigate the influence of the number of positive ion sources on the propulsion developed by an asymmetric capacitor which generates an electrohydrodynamic (EHD) flow through a corona discharge in nitrogen gas, at atmospheric pressure. We are going to study and compare the propulsion efficiencies of three different setups. The only variable between them is the number of corona wires used above the ground electrode, which is the same throughout. The first structure to be studied is an asymmetric capacitor with only one corona wire centered at (0 m, 0.03 m) above the ground electrode with forward section centered at (0 m, 0 m), as can be seen in figure 1. The positive corona wire has always a radius of 0.25 μm, which is a much smaller radius of curvature than the facing ground electrode, with an ellipsoidal cross-section 0.04 m wide and 0.01 m in height. The second structure to be studied is an asymmetric capacitor with two corona wires above the ground electrode which are centered respectively at (-0.005 m, 0.03 m) and (0.005 m, 0.03 m), and is represented in figure 2. The third structure to be studied is a composite of the first two, with three corona wires above the ground electrode, centered at (-0.005 m, 0.03 m), (0 m, 0.03 m) and (0.005 m, 0.03 m), and is represented in figure 3. In the ensuing discussion we are going to apply the known theory of EHD and electrostatics in order to provide a physical insight on the force mechanism that acts on each asymmetrical capacitor, and compare the developed forces in order to determine which is more efficient, or if using a greater number of ion sources augments the propulsion efficiency or not. We will show that these electrostatic actuators generate a propulsion force which differs from the usual plasma actuator (Roth, 2003; Roy, 2005; Pinheiro and Martins, 2010) because they harness the electrostatic force to move the actuator itself.

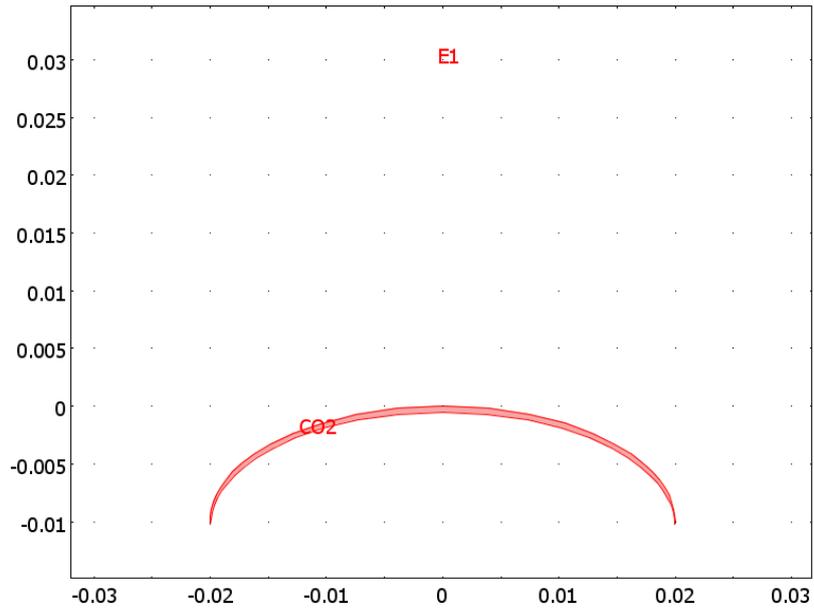

**Figure 1.** Asymmetric capacitor with the corona wire at (0 m, 0.03 m) and forward section of ground electrode centered at (0 m, 0 m).

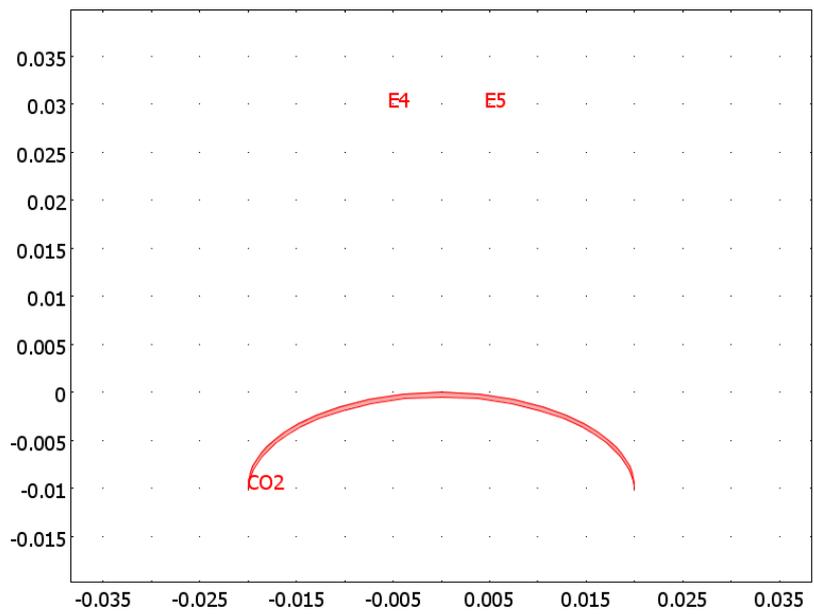

**Figure 2.** Asymmetric capacitor with two corona wires, centered at (-0.005 m, 0.03 m) and (0.005 m, 0.03 m), with forward section of ground electrode centered at (0 m, 0 m).

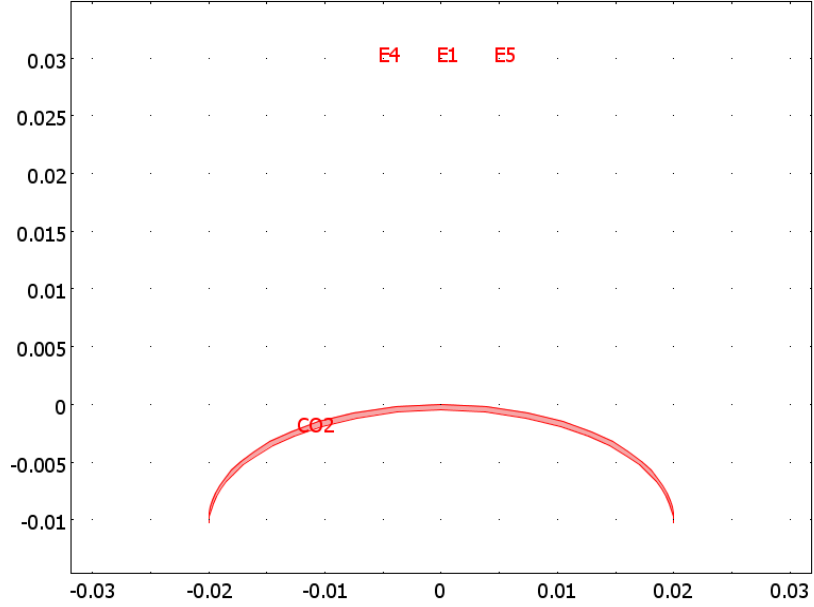

**Figure 3.** Asymmetric capacitor with three corona wires, centered at (-0.005 m, 0.03 m), (0 m, 0.03 m) and (0.005 m, 0.03 m), with forward section of ground electrode centered at (0 m, 0 m).

## 2. Numeric Model

In our numerical model we will consider electrostatic force interactions between the electrodes and the ion space cloud, the moment exchange between the mechanical setup and the induced opposite direction nitrogen flow (EHD flow), nitrogen pressure forces on the structure and viscous drag forces. EHD flow is the flow of neutral particles caused by the drifting of ions in an electric field. In our case these ions are generated by a positive high voltage corona discharge in the high curvature (the higher the radius of a sphere, the less is its curvature) portions of the electrodes. The corona wire has a uniform high curvature and therefore will generate ions uniformly (Chen and Davidson, 2002). On the contrary, the body electrode (ground) has a non uniform curvature having the lowest curvature facing the positive corona wire. The positively ionized gas molecules will travel from the corona wire ion source towards the collector (ground) colliding with neutral molecules in the process. These collisions will impart momentum to the neutral atoms that will move towards the collector as a result. The momentum gained by the neutral gas is exactly equal to the momentum gained by the positive ions accelerated through the electric field between the electrodes, and lost in inelastic collisions to the neutral molecules.

Corona discharges are non-equilibrium plasmas with an extremely low degree of ionization (roughly $10^{-8}$ %). There exist two zones with different properties, the ionization zone and the drift zone. The energy from the electric field is mainly absorbed by electrons in the ionization zone, immediately close to the corona electrode, and transmitted to the neutral gas molecules by inelastic collisions producing electron-positive ion pairs, where the net space charge density $\rho_q$ will remain approximately zero ($\rho_q = 0$). However, the local volume space charge, in the drift zone, will be positive for a positive corona (and therefore $\rho_q = eZ_i n_i$ in the drift zone; $e$ is absolute value of the electron charge, $Z_i$ is the charge of the positive ionic species, $n_i$ is the positive ion density of $N_2^+$) because of the much higher mobility of electrons relative to the positive ions. We will only consider nitrogen ($N_2^+$) positive ions in the drift zone. We make this choice because, for corona discharges in air, the $N_2^+$ and $O_2^+$ ions dominate (Chen and Davidson, 2002), and we want to relate our study with experiments made in an air mixture were nitrogen represents 78% of the total atmospheric gases (22% is Oxygen).

The ionic mobility ($\mu_i$) is defined as the velocity $v$ attained by an ion moving through a gas under unit electric field E (m$^2$ s$^{-1}$ V$^{-1}$), i.e., it is the ratio of the ion drift velocity to the electric field strength:

$$\mu_i = \frac{v}{E}. \quad (1)$$

The mobility is usually a function of the reduced electric field $E/N$ and $T$, where $E$ is the field strength, $N$ is the Loschmidt constant (number of molecules m$^{-3}$ at s.t.p.), and $T$ is the temperature. The unit of $E/N$ is the Townsend (Td), 1 Td = $10^{-21}$ V m$^2$. Since we are applying 28000 V to the corona wire across a gap of 3 cm towards the ground electrode, the reduced electric field will be approximately 38 Td or 38 x 10$^{-17}$ Vcm$^2$ (considering that the gas density N at 1 atm, with a gas temperature $T_g$ of 300 K is N=2,447 x 10$^{19}$ cm$^{-3}$). According to Moseley (1969), the mobility $\mu_i$ of an ion is defined by:

$$\mu_i = \mu_{i0}(760/p)(T/273.16), \quad (2)$$

where $\mu_{i0}$ is the reduced mobility, $p$ is the gas pressure in Torr (1 atm = 760 Torr) and $T$ is the gas temperature in Kelvin. For our experimental condition of E/N = 38 Td, Moseley's measurements indicate a $\mu_{i0}$ of 1,83 cm$^2$/(Vs). Thus, at our operating temperature of 300 K, the mobility $\mu_i$ will be 2,01 cm$^2$/(Vs) or 2,01 x 10$^{-4}$ m$^2$/(Vs).

Since the reduced electric field is relatively low, the ion diffusion coefficient $D_i$ can be approximated by the Einstein relation:

$$D_i = \mu_i \left( \frac{k_B T}{e} \right), \quad (3)$$

where $k_B$ is the Boltzmann constant. This equation provides a diffusion coefficient of 5,19 x 10$^{-6}$ m$^2$/s for our conditions.

The governing equations for EHD flow in an electrostatic fluid accelerator (EFA) are already known (Rickard and Dunn, 2007; Zhao and Adamiak, 2005; Matéo-Vélez et al, 2005) and described next; these will be applied to the drift zone only. The electric field **E** is given by:

$$\mathbf{E} = -\nabla V. \quad (4)$$

Since $\nabla \cdot \mathbf{E} = \frac{\rho_q}{\varepsilon_0}$ (Gauss's law), the electric potential **V** is obtained by solving the Poisson equation:

$$\nabla^2 V = -\frac{\rho_q}{\varepsilon_0} = -\frac{e(Z_i n_i - n_e)}{\varepsilon_0}, \quad (5)$$

where $n_e$ is the negative ion density (we are only considering electrons) and $\varepsilon_0$ is the permittivity of free space. The total volume ionic current density $\mathbf{J}_i$ created by the space charge drift is given by:

$$\mathbf{J}_i = \rho_q \mu_i \mathbf{E} + \rho_q \mathbf{u} - D_i \nabla \rho_q, \quad (6)$$

where $\mu_i$ is the mobility of ions in the nitrogen gas subject to an electric field, $u$ is the gas (nitrogen neutrals) velocity and $D_i$ is the ion diffusion coefficient. The current density satisfies the charge conservation (continuity) equation:

$$\frac{\partial \rho_q}{\partial t} + \nabla \cdot \mathbf{J}_i = 0. \tag{7}$$

But, since we are studying a DC problem, in steady state conditions we have:

$$\nabla \cdot \mathbf{J}_i = 0. \tag{8}$$

The hydrodynamic mass continuity equation for the nitrogen neutrals is given by:

$$\frac{\partial \rho_f}{\partial t} + \nabla \cdot (\rho_f \mathbf{u}) = 0 \tag{9}$$

If the nitrogen fluid density $\rho_f$ is constant, like in incompressible fluids, then it reduces to:

$$\nabla \cdot \mathbf{u} = 0. \tag{10}$$

In this case, the nitrogen is incompressible and it must satisfy the Navier-Stokes equation:

$$\rho_f \left( \frac{\partial \mathbf{u}}{\partial t} + (\mathbf{u} \cdot \nabla) \mathbf{u} \right) = -\nabla p + \mu \nabla^2 \mathbf{u} + \mathbf{f}. \tag{11}$$

The term on the left is considered to be that of inertia, where the first term in brackets is the unsteady acceleration, the second term is the convective acceleration and $\rho_f$ is the density of the hydrodynamic fluid - nitrogen in our case. On the right, the first term is the pressure gradient, the second is the viscosity ($\mu$) force and the third is ascribed to any other external force **f** on the fluid. Since the discharge is DC, the electrical force density on the nitrogen ions that is transferred to the neutral gas is $\mathbf{f}^{EM} = \rho_q \mathbf{E} = -\rho_q \nabla V$. If we insert the current density definition (Equation (6)) into the current continuity (Equation (8)), we obtain the charge transport equation:

$$\nabla \cdot \mathbf{J}_i = \nabla \cdot (\rho_q \mu_i \mathbf{E} + \rho_q \mathbf{u} - D_i \nabla \rho_q) = 0. \tag{12}$$

Since the fluid is incompressible ($\nabla \cdot \mathbf{u} = 0$) this reduces to:

$$\nabla \cdot (\rho_q \mu_i \mathbf{E} - D_i \nabla \rho_q) + \mathbf{u} \nabla \rho_q = 0. \tag{13}$$

In our simulation we will consider all terms present in Equation (13), although it is known that the conduction term (first to the left) is preponderant over the other two (diffusion and convection), since generally the gas velocity is two orders of magnitude smaller than the velocity of ions. Usually, the expression for the current density (Equation (6)) is simplified as:

$$\mathbf{J}_i = \rho_q \mu_i \mathbf{E}, \tag{14}$$

Then, if we insert Equation (14) into Equation (8), expand the divergence and use Equation (4) and Gauss's law we obtain the following (known) equation that describes the evolution of the charge density in the drift zone:

$$\nabla \rho_q \cdot \nabla V - \frac{\rho_q^2}{\varepsilon_0} = 0 \qquad (15)$$

In Table I we can see the values of the parameters used for the simulation. We will consider in our model that the ionization region has zero thickness, as suggested by Morrow (1997). The following equations will be applied to the ionization zone only. For the formulation of the proper boundary conditions for the external surface of the space charge density we will use the Kaptsov hypothesis (Kaptsov, 1947) which states that below corona initiation the electric field and ionization radius will increase in direct proportion to the applied voltage, but will be maintained at a constant value after the corona is initiated.

In our case, a positive space charge $\rho_q$ is generated by the corona wire and drifts towards the ground electrode through the gap $G$ (drift zone) between both electrodes and is accelerated by the local electric field. When the radius of the corona wire is much smaller than $G$, then the ionization zone around the corona wire is uniform. In a positive corona, Peek's empirical formula (Peek, 1929; Cobine, 1958; Meroth, 1999; Atten, Adamiak, and Atrazhev, 2002; Zhao, and Adamiak, 2008) in air gives the electric field strength $E_p$ (V/m) at the surface of an ideally smooth cylindrical wire corona electrode of radius $r_c$:

$$E_p = E_0 \cdot \delta \cdot \varepsilon (1 + 0.308 / \sqrt{\delta \cdot r_c}). \qquad (16)$$

Where $E_0 = 3.31 \cdot 10^6 V/m$ is the dielectric breakdown strength of air, $\delta$ is the relative atmospheric density factor, given by $\delta = 298p/T$, where $T$ is the gas temperature in Kelvin and $p$ is the gas pressure in atmospheres ($T$=300K and $p$=1atm in our model); $\varepsilon$ is the dimensionless surface roughness of the electrode ($\varepsilon = 1$ for a smooth surface) and $r_c$ is given in centimeters. At the boundary between the ionization and drifting zones the electric field strength is equal to $E_0$ according to the Kaptsov assumption. This formula (Peek's law) determines the threshold strength of the electric field to start the corona discharge at the corona wire. Surface charge density will then be calculated by specifying the applied electric potential $V$ and assuming the electric field $E_p$ at the surface of the corona wire. The assumption that the electric field strength at the wire is equal to $E_p$ is justified and discussed by Morrow (1997). Although $E_p$ remains constant after corona initiation, the space charge current $J_i$ will increase with the applied potential $V_c$ in order to keep the electric field at the surface of the corona electrode at the same Peek's value, leading to the increase of the surrounding space charge density and respective radial drift.

Atten, Adamiak and Atrazhev (2002), have compared Peek's empirical formula with other methods including the direct Townsend ionization criterion and despite some differences in the electric field, they concluded that the total corona current differs only slightly for small corona currents (below 6 kV). For voltages above 6 kV (corresponding to higher space currents) the difference is smaller than 10% in the worst case, according to them.

For relatively low space charge density in DC coronas, the electric field $E(r)$ in the plasma (ionization zone) has the form (Chen and Davidson, 2002):

$$E(r) = \frac{E_p r_c}{r}, \qquad (17)$$

where $r$ is the radial position from the center of the corona wire. Since the electric field $E_0$ establishes the frontier to the drift zone, using this formula we can calculate the radius of the ionization zone ($r_i$), which gives:

$$r_i = \frac{E_p r_c}{E_0} = r_c \cdot \delta \cdot \varepsilon (1 + 0.308 / \sqrt{\delta \cdot r_c}). \tag{18}$$

Since we have chosen in our simulation for $r_c$ to be 0,025 *mm*, then $r_i$ would be 0,074 *mm*. Now we can calculate the voltage ($V_i$) at the boundary of the ionization zone by integrating the electric field between $r_c$ and $r_i$:

$$V_i = V_c - E_p r_c \ln(E_p / E_0), \tag{19}$$

where $V_c$ is the voltage applied to the corona electrode and $r_c$ is in meters. This equation is valid only for the ionization zone. In our case it determines that if we apply 28000 Volts to the corona wire, then the voltage present at the boundary of the ionization zone becomes 26833.68 Volts.

For the drift zone, Poisson equation (Equation (5)) should be used together with the charge transport equation (Equation (13)) in order to obtain steady state field and charge density distributions. The values of the relevant parameters for the simulation are detailed in Table I.

Three application modes of the COMSOL 3.5 Multiphysics software are used. The steady state incompressible Navier-Stokes mode is used to resolve the fluid dynamic equations. The electrostatics mode is used to resolve the electric potential distribution and the electrostatic forces to which the electrodes are subjected. The PDE (coefficient form) mode is used to resolve the charge transport equation (Equation (13)). The parameters used for the simulation are shown in Table I. The solution domain was a square of 0.2 m on each side containing the asymmetric capacitor at the center (figure 4).

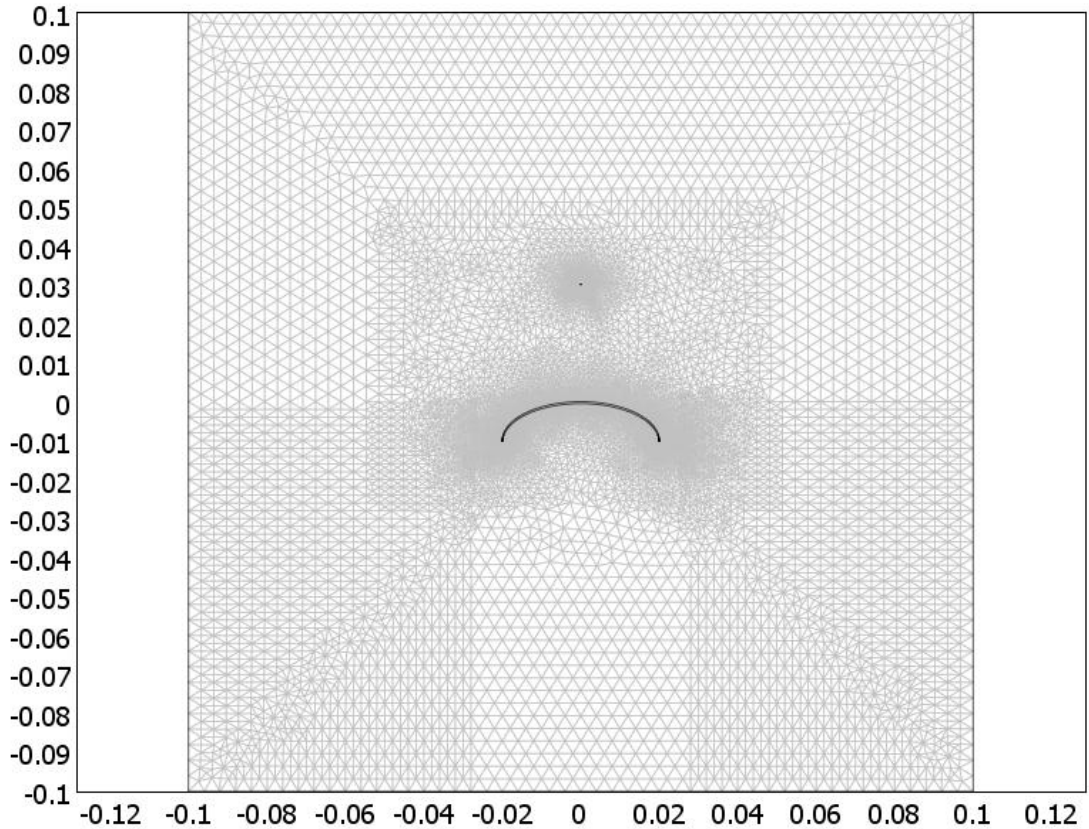

**Figure 4.** Typical mesh of the solution domain (0.2 m × 0.2 m) containing 44270 elements.

**Table I. Value of parameters used for the simulation.**

| Parameters | Value |
|---|---|
| Nitrogen density (T=300K, p=1atm), $\rho_N$ | 1.165 kg/m$^3$ |
| Dynamic viscosity of nitrogen (T=300K, p=1atm), $\mu_N$ | 1,775 x 10$^{-5}$ Ns/m$^2$ |
| Nitrogen relative dielectric permittivity, $\varepsilon_r$ | 1 |
| $N_2^+$ mobility coefficient, $\mu_i$ (for E/N = 54 Td) | 1,92 x 10$^{-4}$ m$^2$/(Vs) |
| $N_2^+$ diffusion coefficient, $D_i$ (for E/N = 54 Td) | 4,96 x 10$^{-6}$ m$^2$/s |
| Corona wire radius, $r_c$ | 25 μm |
| Facing ground electrode width | 0.04 m |
| Ground electrode height | 0.01 m |
| Air gap length | 0.03 m |
| Corona wire voltage, $V_c$ | 28000 V |
| Ground electrode voltage, $V_g$ | 0 V |

## 3. Numerical Simulation Results

All the forces along the vertical (y axis) of the first geometry are presented in Table II. The results of the simulation show that the electrostatic forces $F_{ey}$ on the electrodes are the most relevant forces to consider, constituting 94.92% of the total force. The total hydrodynamic force $F_{HTy}$ is small because the pressure $F_{py}$ and viscosity $F_{vy}$ forces do not contribute in a relevant way in the present conditions. The total resultant force $F_{Ty}$ that acts on the capacitor is -0.324 N/m directed upwards.

**Table II. Forces along the y-axis of the asymmetric capacitor with one wire.**

|  | $F_{py}$ (N/m) | $F_{vy}$ (N/m) | $F_{HTy}$ (N/m) | $F_{ey}$ (N/m) | $F_{Ty}$ (N/m) |
|---|---|---|---|---|---|
| **Corona wire** | -1.453×10$^{-4}$ | 9.199×10$^{-5}$ | -5.333×10$^{-5}$ | -6.193×10$^{-4}$ | -6.726×10$^{-4}$ |
| **Ground electrode** | 1.548×10$^{-2}$ | 1.020×10$^{-3}$ | 1.650×10$^{-2}$ | 3.079×10$^{-1}$ | 3.244×10$^{-1}$ |
| **Total force** | 1.533×10$^{-2}$ | 1.112×10$^{-3}$ | 1.645×10$^{-2}$ | 3.073×10$^{-1}$ | 3.237×10$^{-1}$ |

The corona wire generates a positive charge cloud which accelerates trough the air gap towards the facing ground electrode. The interaction between both electrodes and the positive charge cloud will accelerate the nitrogen positive ions towards the ground electrode and the ions will transmit their momentum to the neutral nitrogen particles by a collision process. Thus, the neutral nitrogen will move in the direction from the corona wire to the ground electrode, and the momentum transmitted to the ions and neutrals is equivalent to the electrostatic forces the ion cloud will induce on both electrodes. The nitrogen neutral wind generated by the collisions with the accelerated ion cloud is shown in figure 5, were the top velocity achieved is 4.295 m/s. The corona wire and nitrogen positive space charge will induce a charge separation in the ground electrode, pulling the electrons or negative charge towards the upper section of the ground electrode which will be subjected to a strong electrostatic force. In this way, the real force that makes the asymmetrical capacitor move is not a moment reaction to the induced air draft but moment reaction to the positive ion cloud that causes the air movement. Total momentum is conserved always. The electrostatic force computation on both electrodes is shown in figure 6.

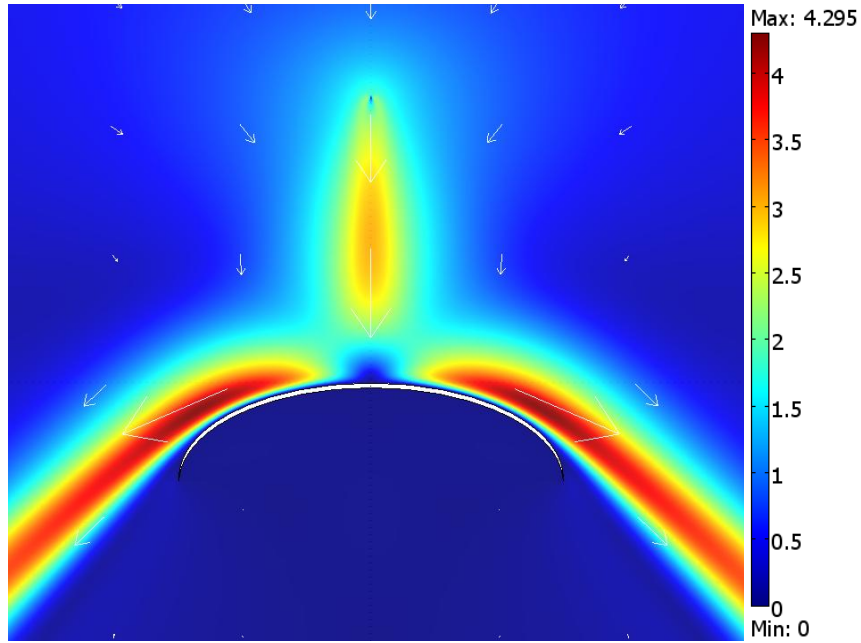

**Figure 5.** Air velocity, for the case with one wire, as surface map with units in m/s with proportional vector arrows.

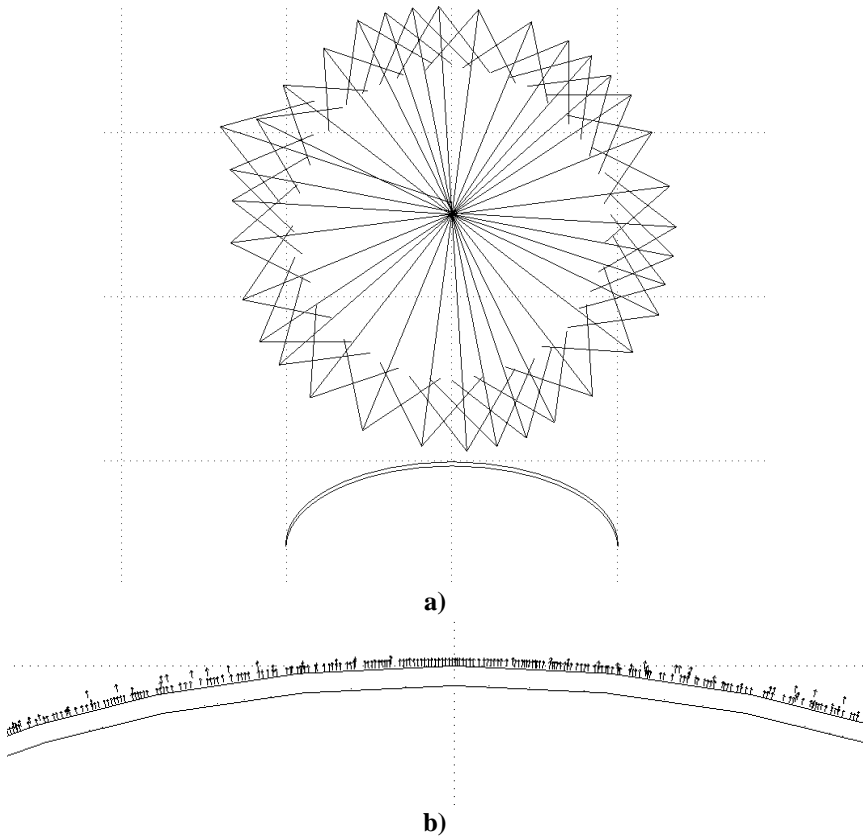

**a)**

**b)**

**Figure 6. a)** Electrostatic force on the corona wire, **b)** Electrostatic force on the ground electrode.

The electrostatic force on the corona wire is very strong but mostly symmetric (figure 6.a)). Nevertheless, there is a slight asymmetry in the positive ion cloud distribution around the corona wire (figure 7) in the direction of the ground electrode. The corona wire constantly creates positive ions, which are strongly attracted towards the ground electrode, creating the asymmetrical distribution. Therefore the electrostatic force on the corona wire will be small and directed towards the ground electrode, due to the positive ion distribution around it. On the other hand some positive ions are neutralized on the ground electrode (figure 6.b)), which requires the consumption of current (electrons) to compensate for the acquired positive charge in order to remain neutral. As shown by Canning, Melcher, and Winet (2004), the current from the power source to the ion emitter is always larger than the current to the ground electrode. This is because not all positive ions emitted by the corona wire will be neutralized in the ground electrode; there may be other neutralizing paths. The electrons that are provided to the neutral electrode (for it to remain neutral) are attracted to and neutralized in the front, were they suffer an electrostatic attraction towards the approaching positive ion cloud (figure 6.b)), subjecting the ground electrode to a strong electrostatic force towards the corona wire. The electrostatic force on the wire is much stronger than that on the ground electrode, but since it is mostly symmetric around the wire the main electrostatic force will be on the ground electrode. Therefore, the ion wind is a reaction to the electrostatic thrust mechanism and not the cause of the thrusting force as it is usually conceived. The main thrust force on the electrodes is electrostatic, not hydrodynamic. Nevertheless, this force mechanism is still dependent on the availability of surrounding particles susceptible to ionization in order to be able to function.

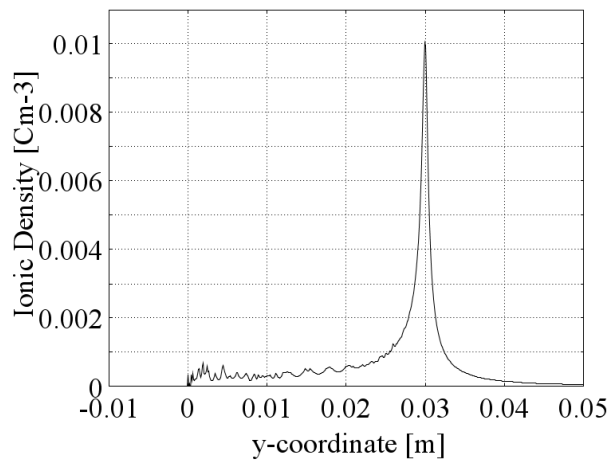

**Figure 7.** Ionic density distribution, for the case with one wire, from (0 m, -0.01 m) to (0 m, 0.05 m).

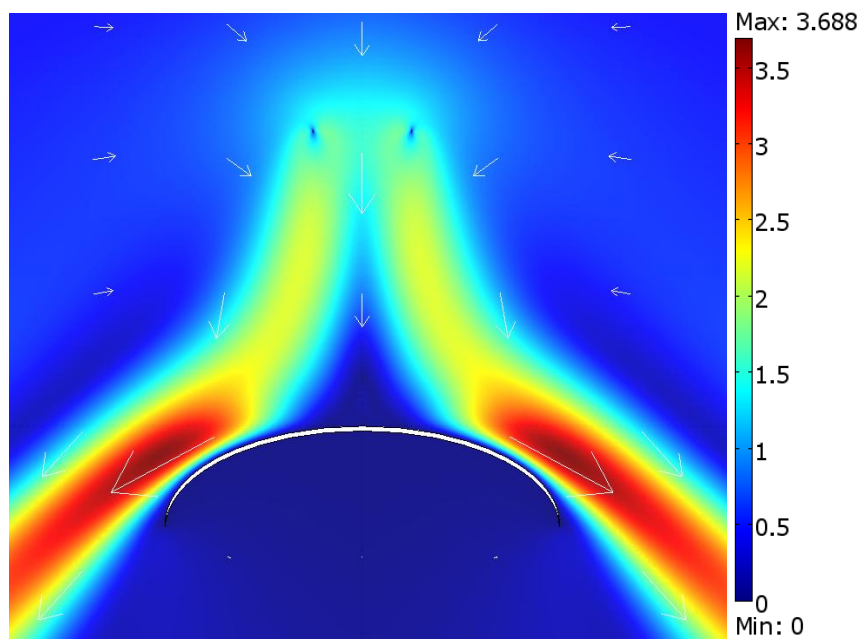

**Figure 8.** Air velocity, for the case with two wires, as surface map with units in m/s with proportional vector arrows.

For the asymmetrical capacitor with two wires, the highest air velocity achieved is 3.688 m/s as shown on figure 8. If we analyze the lines of equal ionic density for this case (figure 9) we can clearly see the neutralization of the positive ions on the ground electrode, and the ion charge asymmetry around the corona wires due to the electric attraction towards the ground electrode. The total hydrodynamic and electrostatic forces for this case are presented in Table III.

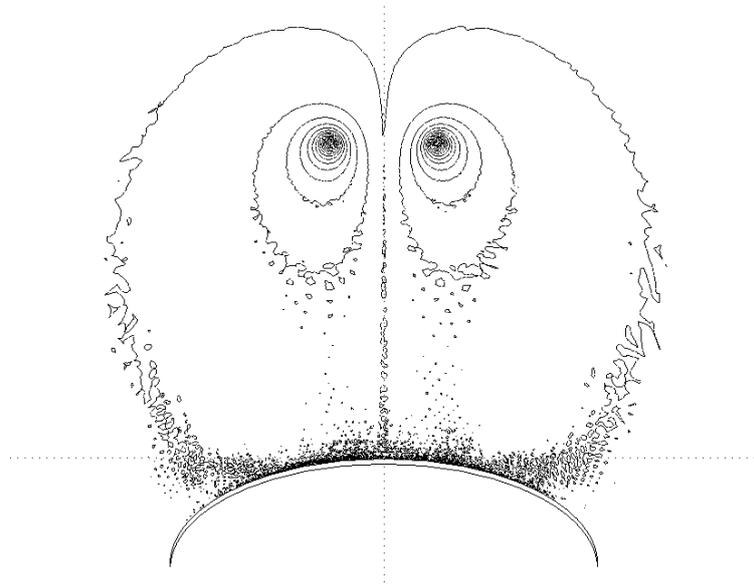

**Figure 9.** Lines of equal ionic density, for the case with two wires.

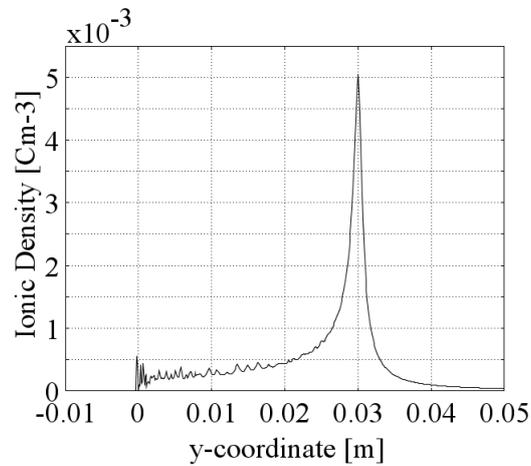

**Figure 10.** Ionic density distribution [$Cm^{-3}$], for the case with two wires, from (-0.005 m, -0.01 m) to (-0.005 m, 0.05 m).

**Table III. Forces along the y-axis of the asymmetric capacitor with two wires.**

|  | $F_{py}$ (N/m) | $F_{vy}$ (N/m) | $F_{HTy}$ (N/m) | $F_{ey}$ (N/m) | $F_{Ty}$ (N/m) |
|---|---|---|---|---|---|
| **Left Corona wire** | $-2.789 \times 10^{-5}$ | $5.095 \times 10^{-5}$ | $2.306 \times 10^{-5}$ | $-8.079 \times 10^{-4}$ | $-7.848 \times 10^{-4}$ |
| **Right Corona wire** | $-2.795 \times 10^{-5}$ | $5.123 \times 10^{-5}$ | $2.327 \times 10^{-5}$ | $-8.239 \times 10^{-4}$ | $-8.006 \times 10^{-4}$ |
| **Ground electrode** | $5.183 \times 10^{-3}$ | $5.882 \times 10^{-4}$ | $5.771 \times 10^{-3}$ | $3.475 \times 10^{-1}$ | $3.533 \times 10^{-1}$ |
| **Total force** | $5.127 \times 10^{-3}$ | $6.903 \times 10^{-4}$ | $5.817 \times 10^{-3}$ | $3.459 \times 10^{-1}$ | $3.517 \times 10^{-1}$ |

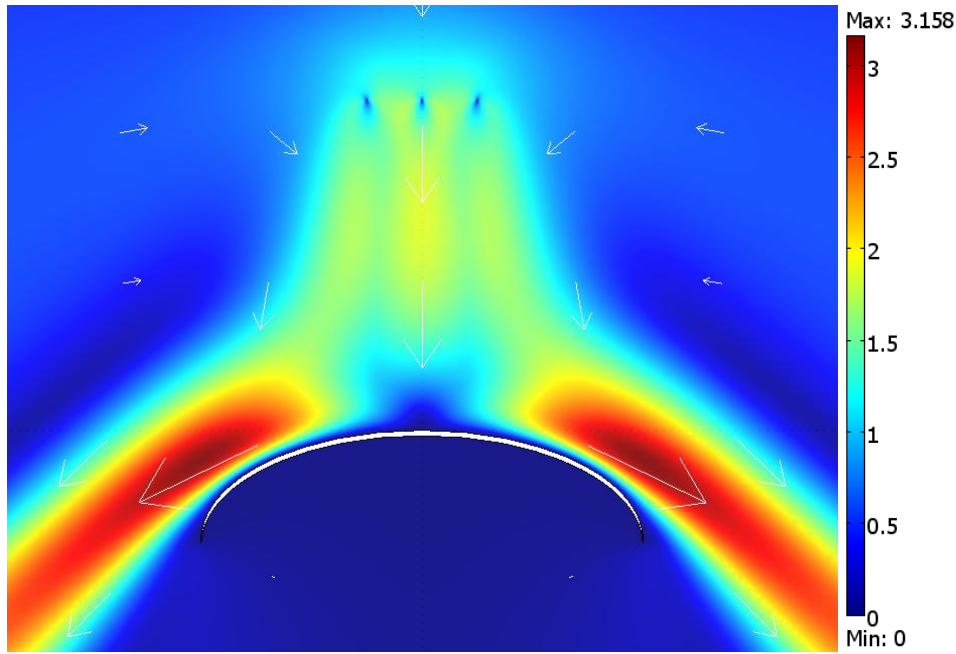

**Figure 11.** Air velocity, for the case with three wires, as surface map with units in m/s with proportional vector arrows.

If the asymmetrical capacitor has three wires, the highest air velocity achieved is 3.158 m/s as shown on figure 11, and the total hydrodynamic and electrostatic forces for this case are presented in Table IV. Comparing the total forces for each case (Table V) we can clearly see that the highest propulsive force occurs for the capacitor with two wires, therefore the addition of a third wire is not beneficial. The main force responsible for the propulsion in all cases is electrostatic and the total force is much higher than the force (0.182 N/m) produced by the lifter geometry (Martins and Pinheiro, 2010) due to the increase of the ground electrode surface exposed to electrostatic attraction.

**Table IV. Forces along the y-axis of the asymmetric capacitor with three wires.**

|  | $F_{py}$ (N/m) | $F_{vy}$ (N/m) | $F_{HTy}$ (N/m) | $F_{ey}$ (N/m) | $F_{Ty}$ (N/m) |
|---|---|---|---|---|---|
| **Left Corona wire** | $-2.195 \times 10^{-5}$ | $-9.058 \times 10^{-7}$ | $-2.286 \times 10^{-5}$ | $-5.049 \times 10^{-3}$ | $-5.072 \times 10^{-3}$ |
| **Center Corona wire** | $-2.130 \times 10^{-5}$ | $-1.386 \times 10^{-6}$ | $-2.269 \times 10^{-5}$ | $-3.914 \times 10^{-3}$ | $-3.937 \times 10^{-3}$ |
| **Right Corona wire** | $-2.265 \times 10^{-5}$ | $-1.297 \times 10^{-6}$ | $-2.395 \times 10^{-5}$ | $-4.852 \times 10^{-3}$ | $-4.876 \times 10^{-3}$ |
| **Ground electrode** | $1.548 \times 10^{-2}$ | $1.020 \times 10^{-3}$ | $1.650 \times 10^{-2}$ | $2.911 \times 10^{-1}$ | $3.076 \times 10^{-1}$ |
| **Total force** | $1.541 \times 10^{-2}$ | $1.016 \times 10^{-3}$ | $1.643 \times 10^{-2}$ | $2.773 \times 10^{-1}$ | $2.937 \times 10^{-1}$ |

**Table V. Comparison of the total force on the considered asymmetric capacitors.**

|  | $F_{Ty}$ (N/m) | % $F_{ey}$ |
|---|---|---|
| **One wire** | 0.3237 | 94.92 |
| **Two wires** | 0.3517 | 98.35 |
| **Three wires** | 0.2937 | 94.41 |

## 4. CONCLUSION

The physical origin of the force that acts on an asymmetrical capacitor is electrostatic, with a component always superior to 94.4% on all cases, mainly concentrated on the ground electrode. The electrostatic force vectors on the wire are much stronger than that on the ground electrode, but since it is mostly symmetric around the wire the main electrostatic force will be on the ground electrode, where the electrostatic force is much smaller but applied to a much larger surface and therefore significant. The generated ion wind is a reaction to the electrostatic thrust mechanism and not the cause of the thrusting force as it is usually conceived. The main thrust force on the electrodes is electrostatic, not hydrodynamic. Increasing the number of wires will not necessarily increase the generated force. In fact the highest air velocity decreases with the increase of the number of wires, most probably due to the electric field decrease in each wire due to the interaction between neighboring wires. The most efficient configuration is with two corona wires, where the electrostatic force component is highest, perhaps due to the induction of a uniform and wider distribution of positive ions above the surface of the ground electrode.

**Acknowledgements**

The authors gratefully thank to Mário Lino da Silva for the permission to use his computer with 32 gigabytes of RAM and two quad-core processors, without which this work would not have been possible. We also acknowledge partial financial support by the Reitoria da Universidade Técnica de Lisboa. We would like to thank important financial support to one of the authors, AAM, in the form of a PhD Scholarship from FCT (Fundação para a Ciência e a Tecnologia).